\documentstyle[10pt,epsf,epsfig,dp_delphititle]{dp_delphi}
\makeindex
\def\DpPaperGroup{EP}
\def\DpPaperRef{2000-010}
\def\DpDate{18 January 2000}
\def\DpAuthors{DELPHI Collaboration}
\def\DpSubmit{(Phys. Lett. B479(2000)118; erratum Phys. Lett. B492(2000)398)}
\def\DpTitle{{Hadronization properties of b quarks compared to light 
        quarks\\in \boldmath $e^+e^- \rightarrow q\bar{q}$ from 
        $183$ to $200$ GeV}}
\def\DpComment{ }
\def\DpEMail{ }
\def\bb{{b\bar{b}}}
\def\cc{{c\bar{c}}}
\def\ll{{l\bar{l}}}
\def\lbl{{l\bar{l}}}
\newcommand{\kos}{\ifmmode {{\mathrm K}^{0}_{S} } \else
${\mathrm K}^{0}_{S}$ \fi}
\newcommand{\kpm}{\ifmmode {{\mathrm K}^{\pm}} \else
${\mathrm K}^{\pm}$\fi}
\newcommand{\ko}{\ifmmode {{\mathrm K}^{0}} \else
${\mathrm K}^{0}$ \fi}

\def\ZP{Z.\ Phys.\ {\bf C}}
\def\PL{Phys.\ Lett.\ {\bf B}}

\def\PRL{Phys.\ Rev.\ Lett.\ }

\def\NIM{Nucl.\ Instr.\ Meth.\ }

\def\Abreu{DELPHI Coll., P. Abreu {et al.,}\ }

\begin{document}
%%%%%%%%%%%%%%%%%%%%%%%%%% They are a problem with Coll.Sty ?
\makeatletter
%\input{dp_system:coll.sty}
% Collapse citation numbers to ranges.  Non-numeric and undefined labels
% are handled.  No sorting is done.  E.g., 1,3,2,3,4,5,foo,1,2,3,?,4,5
% gives 1,3,2-5,foo,1-3,?,4,5
\newcount\@tempcntc
\def\@citex[#1]#2{\if@filesw\immediate\write\@auxout{\string\citation{#2}}\fi
  \@tempcnta\z@\@tempcntb\m@ne\def\@citea{}\@cite{\@for\@citeb:=#2\do
    {\@ifundefined
       {b@\@citeb}{\@citeo\@tempcntb\m@ne\@citea\def\@citea{,}{\bf ?}\@warning
       {Citation `\@citeb' on page \thepage \space undefined}}%
    {\setbox\z@\hbox{\global\@tempcntc0\csname b@\@citeb\endcsname\relax}%
     \ifnum\@tempcntc=\z@ \@citeo\@tempcntb\m@ne
       \@citea\def\@citea{,}\hbox{\csname b@\@citeb\endcsname}%
     \else
      \advance\@tempcntb\@ne
      \ifnum\@tempcntb=\@tempcntc
      \else\advance\@tempcntb\m@ne\@citeo
      \@tempcnta\@tempcntc\@tempcntb\@tempcntc\fi\fi}}\@citeo}{#1}}
\def\@citeo{\ifnum\@tempcnta>\@tempcntb\else\@citea\def\@citea{,}%
  \ifnum\@tempcnta=\@tempcntb\the\@tempcnta\else
   {\advance\@tempcnta\@ne\ifnum\@tempcnta=\@tempcntb \else \def\@citea{--}\fi
    \advance\@tempcnta\m@ne\the\@tempcnta\@citea\the\@tempcntb}\fi\fi}
 
\makeatother
%%%%%%%%%%%%%%%%%%%%%%%%%% ??????????????????????????????????
% Generate the title page
\begin{titlepage}
\pagenumbering{roman}
\CERNpreprint{\DpPaperGroup}{\DpPaperRef} % Reference of the paper
\date{{\small\DpDate}} % Date of the paper
\title{\DpTitle} % Title of the paper
\address{\DpAuthors} % General name of the author(s)
\begin{shortabs} % Start the abstract
\noindent
%   abstract.tex
%
\noindent

The DELPHI detector at LEP has collected 54 pb$^{-1}$ of data
at a centre-of-mass energy around 183 GeV during 1997,
158 pb$^{-1}$ around 189 GeV during 1998,
and 187 pb$^{-1}$ between 192 and 200 GeV during 1999. 
These data were used to
measure the average charged particle multiplicity in $e^+e^-
\rightarrow b\bar{b}$ events, $\langle n\rangle_\bb$, and the difference $\delta_{bl}$ 
between $\langle n\rangle_\bb$ and the
multiplicity, $\langle n\rangle_\ll$, in generic light quark (u,d,s) events:
\begin{eqnarray*}
\delta_{bl}\rm{(183 \, GeV)} &=& 4.55 \pm 1.31 (stat) \pm 0.73 (syst) \\
\delta_{bl}\rm{(189 \, GeV)} &=& 4.43 \pm 0.85 (stat) \pm 0.61 (syst) \\
\delta_{bl}\rm{(200 \, GeV)} &=& 3.39 \pm 0.89 (stat) \pm 1.01 (syst) \, .
\end{eqnarray*}
This result is consistent with QCD predictions, while it is
inconsistent with calculations assuming that the 
multiplicity
accompanying the decay of a heavy quark is independent of the mass of
the quark itself.

\end{shortabs}
\vfill
\begin{center}
\DpSubmit \ \\ % Horrible hack to allow to have DpSubmit empty
\DpComment \ \\
\DpEMail \ \\
\end{center}
\vfill
\clearpage
\headsep 10.0pt
\addtolength{\textheight}{10mm}
\addtolength{\footskip}{-5mm}
\begingroup
% Commands to process the author names
%
\newcommand{\DpName}[2]{\hbox{#1$^{\ref{#2}}$},\hfill}
\newcommand{\DpNameTwo}[3]{\hbox{#1$^{\ref{#2},\ref{#3}}$},\hfill}
\newcommand{\DpNameThree}[4]{\hbox{#1$^{\ref{#2},\ref{#3},\ref{#4}}$},\hfill}
\newskip\Bigfill \Bigfill = 0pt plus 1000fill
\newcommand{\DpNameLast}[2]{\hbox{#1$^{\ref{#2}}$}\hspace{\Bigfill}}
%
%\small
\footnotesize
\noindent
\DpName{P.Abreu}{LIP}
\DpName{W.Adam}{VIENNA}
\DpName{T.Adye}{RAL}
\DpName{P.Adzic}{DEMOKRITOS}
\DpName{Z.Albrecht}{KARLSRUHE}
\DpName{T.Alderweireld}{AIM}
\DpName{G.D.Alekseev}{JINR}
\DpName{R.Alemany}{VALENCIA}
\DpName{T.Allmendinger}{KARLSRUHE}
\DpName{P.P.Allport}{LIVERPOOL}
% ...
\normalsize
\endgroup
\titlefoot{Department of Physics and Astronomy, Iowa State
     University, Ames IA 50011-3160, USA
    \label{AMES}}
\titlefoot{Physics Department, Univ. Instelling Antwerpen,
     Universiteitsplein 1, B-2610 Antwerpen, Belgium \\
     \indent~~and IIHE, ULB-VUB,
     Pleinlaan 2, B-1050 Brussels, Belgium \\
     \indent~~and Facult\'e des Sciences,
     Univ. de l'Etat Mons, Av. Maistriau 19, B-7000 Mons, Belgium
    \label{AIM}}
\titlefoot{Physics Laboratory, University of Athens, Solonos Str.
     104, GR-10680 Athens, Greece
    \label{ATHENS}}
\titlefoot{Department of Physics, University of Bergen,
     All\'egaten 55, NO-5007 Bergen, Norway
    \label{BERGEN}}
\titlefoot{Dipartimento di Fisica, Universit\`a di Bologna and INFN,
     Via Irnerio 46, IT-40126 Bologna, Italy
    \label{BOLOGNA}}
\titlefoot{Centro Brasileiro de Pesquisas F\'{\i}sicas, rua Xavier Sigaud 150,
     BR-22290 Rio de Janeiro, Brazil \\
     \indent~~and Depto. de F\'{\i}sica, Pont. Univ. Cat\'olica,
     C.P. 38071 BR-22453 Rio de Janeiro, Brazil \\
     \indent~~and Inst. de F\'{\i}sica, Univ. Estadual do Rio de Janeiro,
     rua S\~{a}o Francisco Xavier 524, Rio de Janeiro, Brazil
    \label{BRASIL}}
\titlefoot{Comenius University, Faculty of Mathematics and Physics,
     Mlynska Dolina, SK-84215 Bratislava, Slovakia
    \label{BRATISLAVA}}
\titlefoot{Coll\`ege de France, Lab. de Physique Corpusculaire, IN2P3-CNRS,
     FR-75231 Paris Cedex 05, France
    \label{CDF}}
\titlefoot{CERN, CH-1211 Geneva 23, Switzerland
    \label{CERN}}
\titlefoot{Institut de Recherches Subatomiques, IN2P3 - CNRS/ULP - BP20,
     FR-67037 Strasbourg Cedex, France
    \label{CRN}}
\titlefoot{Now at DESY-Zeuthen, Platanenallee 6, D-15735 Zeuthen, Germany
    \label{DESY}}
\titlefoot{Institute of Nuclear Physics, N.C.S.R. Demokritos,
     P.O. Box 60228, GR-15310 Athens, Greece
    \label{DEMOKRITOS}}
\titlefoot{FZU, Inst. of Phys. of the C.A.S. High Energy Physics Division,
     Na Slovance 2, CZ-180 40, Praha 8, Czech Republic
    \label{FZU}}
\titlefoot{Dipartimento di Fisica, Universit\`a di Genova and INFN,
     Via Dodecaneso 33, IT-16146 Genova, Italy
    \label{GENOVA}}
\titlefoot{Institut des Sciences Nucl\'eaires, IN2P3-CNRS, Universit\'e
     de Grenoble 1, FR-38026 Grenoble Cedex, France
    \label{GRENOBLE}}
\titlefoot{Helsinki Institute of Physics, HIP,
     P.O. Box 9, FI-00014 Helsinki, Finland
    \label{HELSINKI}}
\titlefoot{Joint Institute for Nuclear Research, Dubna, Head Post
     Office, P.O. Box 79, RU-101 000 Moscow, Russian Federation
    \label{JINR}}
\titlefoot{Institut f\"ur Experimentelle Kernphysik,
     Universit\"at Karlsruhe, Postfach 6980, DE-76128 Karlsruhe,
     Germany
    \label{KARLSRUHE}}
\titlefoot{Institute of Nuclear Physics and University of Mining and Metalurgy,
     Ul. Kawiory 26a, PL-30055 Krakow, Poland
    \label{KRAKOW}}
\titlefoot{Universit\'e de Paris-Sud, Lab. de l'Acc\'el\'erateur
     Lin\'eaire, IN2P3-CNRS, B\^{a}t. 200, FR-91405 Orsay Cedex, France
    \label{LAL}}
\titlefoot{School of Physics and Chemistry, University of Lancaster,
     Lancaster LA1 4YB, UK
    \label{LANCASTER}}
\titlefoot{LIP, IST, FCUL - Av. Elias Garcia, 14-$1^{o}$,
     PT-1000 Lisboa Codex, Portugal
    \label{LIP}}
\titlefoot{Department of Physics, University of Liverpool, P.O.
     Box 147, Liverpool L69 3BX, UK
    \label{LIVERPOOL}}
\titlefoot{LPNHE, IN2P3-CNRS, Univ.~Paris VI et VII, Tour 33 (RdC),
     4 place Jussieu, FR-75252 Paris Cedex 05, France
    \label{LPNHE}}
\titlefoot{Department of Physics, University of Lund,
     S\"olvegatan 14, SE-223 63 Lund, Sweden
    \label{LUND}}
\titlefoot{Universit\'e Claude Bernard de Lyon, IPNL, IN2P3-CNRS,
     FR-69622 Villeurbanne Cedex, France
    \label{LYON}}
\titlefoot{Univ. d'Aix - Marseille II - CPP, IN2P3-CNRS,
     FR-13288 Marseille Cedex 09, France
    \label{MARSEILLE}}
\titlefoot{Dipartimento di Fisica, Universit\`a di Milano and INFN-MILANO,
     Via Celoria 16, IT-20133 Milan, Italy
    \label{MILANO}}
\titlefoot{Dipartimento di Fisica, Univ. di Milano-Bicocca and
     INFN-MILANO, Piazza delle Scienze 2, IT-20126 Milan, Italy
    \label{MILANO2}}
\titlefoot{Niels Bohr Institute, Blegdamsvej 17,
     DK-2100 Copenhagen {\O}, Denmark
    \label{NBI}}
\titlefoot{IPNP of MFF, Charles Univ., Areal MFF,
     V Holesovickach 2, CZ-180 00, Praha 8, Czech Republic
    \label{NC}}
\titlefoot{NIKHEF, Postbus 41882, NL-1009 DB
     Amsterdam, The Netherlands
    \label{NIKHEF}}
\titlefoot{National Technical University, Physics Department,
     Zografou Campus, GR-15773 Athens, Greece
    \label{NTU-ATHENS}}
\titlefoot{Physics Department, University of Oslo, Blindern,
     NO-1000 Oslo 3, Norway
    \label{OSLO}}
\titlefoot{Dpto. Fisica, Univ. Oviedo, Avda. Calvo Sotelo
     s/n, ES-33007 Oviedo, Spain
    \label{OVIEDO}}
\titlefoot{Department of Physics, University of Oxford,
     Keble Road, Oxford OX1 3RH, UK
    \label{OXFORD}}
\titlefoot{Dipartimento di Fisica, Universit\`a di Padova and
     INFN, Via Marzolo 8, IT-35131 Padua, Italy
    \label{PADOVA}}
\titlefoot{Rutherford Appleton Laboratory, Chilton, Didcot
     OX11 OQX, UK
    \label{RAL}}
\titlefoot{Dipartimento di Fisica, Universit\`a di Roma II and
     INFN, Tor Vergata, IT-00173 Rome, Italy
    \label{ROMA2}}
\titlefoot{Dipartimento di Fisica, Universit\`a di Roma III and
     INFN, Via della Vasca Navale 84, IT-00146 Rome, Italy
    \label{ROMA3}}
\titlefoot{DAPNIA/Service de Physique des Particules,
     CEA-Saclay, FR-91191 Gif-sur-Yvette Cedex, France
    \label{SACLAY}}
\titlefoot{Instituto de Fisica de Cantabria (CSIC-UC), Avda.
     los Castros s/n, ES-39006 Santander, Spain
    \label{SANTANDER}}
\titlefoot{Dipartimento di Fisica, Universit\`a degli Studi di Roma
     La Sapienza, Piazzale Aldo Moro 2, IT-00185 Rome, Italy
    \label{SAPIENZA}}
\titlefoot{Inst. for High Energy Physics, Serpukov
     P.O. Box 35, Protvino, (Moscow Region), Russian Federation
    \label{SERPUKHOV}}
\titlefoot{J. Stefan Institute, Jamova 39, SI-1000 Ljubljana, Slovenia
     and Laboratory for Astroparticle Physics,\\
     \indent~~Nova Gorica Polytechnic, Kostanjeviska 16a, SI-5000 Nova Gorica, Slovenia, \\
     \indent~~and Department of Physics, University of Ljubljana,
     SI-1000 Ljubljana, Slovenia
    \label{SLOVENIJA}}
\titlefoot{Fysikum, Stockholm University,
     Box 6730, SE-113 85 Stockholm, Sweden
    \label{STOCKHOLM}}
\titlefoot{Dipartimento di Fisica Sperimentale, Universit\`a di
     Torino and INFN, Via P. Giuria 1, IT-10125 Turin, Italy
    \label{TORINO}}
\titlefoot{Dipartimento di Fisica, Universit\`a di Trieste and
     INFN, Via A. Valerio 2, IT-34127 Trieste, Italy \\
     \indent~~and Istituto di Fisica, Universit\`a di Udine,
     IT-33100 Udine, Italy
    \label{TU}}
\titlefoot{Univ. Federal do Rio de Janeiro, C.P. 68528
     Cidade Univ., Ilha do Fund\~ao
     BR-21945-970 Rio de Janeiro, Brazil
    \label{UFRJ}}
\titlefoot{Department of Radiation Sciences, University of
     Uppsala, P.O. Box 535, SE-751 21 Uppsala, Sweden
    \label{UPPSALA}}
\titlefoot{IFIC, Valencia-CSIC, and D.F.A.M.N., U. de Valencia,
     Avda. Dr. Moliner 50, ES-46100 Burjassot (Valencia), Spain
    \label{VALENCIA}}
\titlefoot{Institut f\"ur Hochenergiephysik, \"Osterr. Akad.
     d. Wissensch., Nikolsdorfergasse 18, AT-1050 Vienna, Austria
    \label{VIENNA}}
\titlefoot{Inst. Nuclear Studies and University of Warsaw, Ul.
     Hoza 69, PL-00681 Warsaw, Poland
    \label{WARSZAWA}}
\titlefoot{Fachbereich Physik, University of Wuppertal, Postfach
     100 127, DE-42097 Wuppertal, Germany
    \label{WUPPERTAL}}
\addtolength{\textheight}{-10mm}
\addtolength{\footskip}{5mm}
\clearpage
\headsep 30.0pt
\end{titlepage}
%%%%%%%%%%%%%%%%%%%%%%%%%
%
% Change for the document body
%%\pagestyle{heading} % for page numbering
\pagenumbering{arabic} % page numbering in number
\setcounter{footnote}{0} %
\large
%\linenumbers %%%CD
\section{Introduction}

The study of the properties of 
the fragmentation of heavy quarks compared to
light quarks  offers new 
insights in perturbative QCD. Particularly important is
the difference in charged particle multiplicity between 
light quark and heavy quark initiated events in
$e^+e^-$ annihilations. 

In a first approximation one could expect that the multiplicity of hadrons 
produced in addition to the possible decay products of the primary 
quark-antiquark is a universal function of the available
invariant mass; this would 
give a  difference in charged particle multiplicity between 
light quark and heavy quark initiated events decreasing with the
centre-of-mass energy $E_{cm}$ \cite{kisselev}.
QCD predicts, somehow counter-intuitively, 
that this difference is energy independent; this is motivated by
mass effects on the gluon radiation
(see \cite{schumm,petrov,deus} and \cite{khoze} for a recent review).

The existing experimental tests were not conclusive 
(see \cite{schumm} and references therein,
\cite{delphi,opal,sld,tristan}). 
At LEP 2 energies, however, the difference 
between the QCD prediction and the model ignoring mass effects 
is large, and the experimental measurement can firmly distinguish between the
two hypotheses.

\section{Analysis and Results}

A description of the DELPHI detector can be found in \cite{deldet}; its
performance is discussed in \cite{perfo}.

Data corresponding to
a luminosity of 54 pb$^{-1}$ collected by DELPHI 
at centre-of-mass (c.m.) energies around 183~GeV during 1997,
to 158 pb$^{-1}$ collected
around 189~GeV during 1998, and to
187 pb$^{-1}$ collected between 192 and 200 GeV during 1999,
were analysed. 

The 1999 data were taken at different energies: 
25.8 pb$^{-1}$ at 192 GeV, 77.4 pb$^{-1}$ at 196 GeV and 83.8 
pb$^{-1}$ at 200 GeV. Each energy was analyzed separately and the
results were then combined as described later and attributed to 
a c.m. energy of 200 GeV.

A preselection of hadronic events was made, requiring
at least 10
charged particles with momentum $p$ above 100 MeV/$c$ and 
less than 1.5 times the beam energy, with an 
angle $\theta$ with respect to the beam direction
between 20$^\circ$ and 160$^\circ$,
a track length of at least 30~cm,
a distance of closest approach to the interaction point
less than 4~cm in the plane
perpendicular to the beam axis and
less than (4/$\sin\theta$) cm along the beam axis,
a relative error on the momentum measurement $\Delta p/p < 1$,
and a
total transverse energy of the charged particles above 0.2$E_{cm}$.

The influence of the detector on the analysis was studied with
the full DELPHI simulation program, DELSIM~\cite{perfo}.
Events were generated with
PYTHIA 5.7 and JETSET 7.4~\cite{lund},
with parameters tuned to fit LEP1 data from DELPHI \cite{tuning}. The
Parton Shower (PS) model was used.
The particles were followed through
the detailed geometry of DELPHI giving simulated digitisations in each
subdetector. These data were processed with the same
reconstruction and analysis programs as the real data.

The hadronic cross-section for $e^+e^-$ interactions 
above the Z peak is dominated by radiative q$\bar{\mathrm q}\gamma$ events;
the initial state radiated photons (ISR photons) are
generally aligned along the beam direction and not detected.
In order to compute the hadronic
c.m. energy, the procedure described in \cite{sprime} was used.
In this procedure particles are
clustered into jets 
and the effective centre-of-mass energy of the
hadronic system, $\sqrt{s^\prime}$, is computed as being the invariant
mass of the system recoiling against an ISR photon, possibly unseen.

Events with reconstructed hadronic c.m. energy
($\sqrt{s'}$) above 0.9$E_{cm}$ were used.
The selected 1997 (1998, 1999) data sample consisted of 1699 (4583, 4881) 
hadronic events. 

For each year's data, two samples enriched in 
(1) $b-$ events and in (2) $uds-$ events were selected from the
$b$ tagging variable $y$ defined as in Ref. \cite{perfo}; this 
variable represents essentially the probability that none of the
tracks in the event comes from a vertex separated from the primary one.
To select the samples of the type 
(2), it was required in addition that the narrow
jet broadening $B_{min}$ is smaller than 0.065,
to reduce the background due to WW and ZZ events. $B_{min}$ 
is defined as follows.
The event is separated into two hemispheres $H_1$ and $H_2$ with respect to
the thrust axis,
defined by the thrust unit vector $\hat{t}$.
Then, calling $\vec{p_k}$ the momentum vector of the $k$-th particle, 
$$B_{min} = \min_{i=1,2} \frac{\sum_{k \in H_i} |\vec{p_k} \times \hat{t}|}
{2 \sum_k |\vec{p_k}| } \, .$$ 

The contamination from non-q$\bar{\mathrm q}$ events in the samples of
type (1) was 7\% (8\%, 15\%), while it was
13\% (17\%, 20\%) in the samples of type (2).
After applying the event selection
criteria and the cuts to reduce the WW and ZZ background, the  
purities were approximately 91\%
(90\%, 90\%) 
($b-$ events) over the total q$\bar{\mathrm q}$ in sample (1), and 79\%
(79\%, 79\%)  
($uds-$ events)  over the total q$\bar{\mathrm q}$ in sample (2). 
The fractions of $q$-type quarks in the $(i)$-th sample, 
$f^{(i)}_q$, were determined from the simulation. The sample (1) consisted of
103 (326, 416) events; the sample (2) of 590 (1450, 1652) events.
% Coherently with the purities given above, here the numbers of events are
% AFTER BACKGROUND (WW+ZZ) subtracted.
% The numbers bef. backgr. subtr.: 112 (353, 491) (1) and 681 (1753, 2247) (2)

The average charge multiplicity was
measured in the samples (1) and (2), after subtracting the background 
bin-by-bin by means of the simulation. 
It should be noted that the average
multiplicity for a given flavour $q$ in each 
sample is equal to $C\sp{(i)}_q \times \langle n\rangle_{q\bar{q}}$, with  
$C\sp{(i)}_q \neq 1$ in general. The factors $C\sp{(i)}_q$ 
account for biases introduced by the
application of the $b$ probability and the jet broadening cuts, as well
as for detector effects;
these factors were computed by means of the simulation.

A third sample (3)
was taken into account by considering the measurement of multiplicity 
described in \cite{noi}. This 
measurement was performed from a sample of 1297 (3444, 3648) hadronic events, 
with a contamination of 11\%
(14\%, 18\%) 
after applying all the selection criteria; the remaining background 
mostly comes from the hadronic decay of W and Z pairs.
The values $\langle n\rangle^{(3)}$ shown in Table \ref{puri} are fully corrected for
 these backgrounds and for detector effects with their statistical errors;
 hence the nominal quark flavour ratios appear in the equation (3) below.
 The systematic errors are reported as the last contribution in 
Table \ref{sys}.

The measured mean multiplicities together with the event
probability cuts and the factors 
$f^{(i)}_q$ and $C\sp{(i)}_q$ are shown in Table~\ref{puri}.
For the 1999 data, the values only at $\sqrt{s} = 200$ GeV are tabulated.
\begin{table}[htbp]
\begin{center}
\begin{tabular}{|c|c|c|c|c|c|c|c|c|}\hline
\multicolumn{9}{|c|}{Data at 183 GeV}\\ \hline
{\rm Sample} & b-tag prob. & $f\sp{(i)}_b$ & $C\sp{(i)}_b$ & $f\sp{(i)}_{uds}$
& $C\sp{(i)}_{uds}$ & $f\sp{(i)}_c$ & $C\sp{(i)}_c$ & $\langle n\rangle ^{(i)}$ \\ \hline
${\rm (1) }$ & $P_E<0.00001$ & 0.914 & 0.921
& 0.017 & 1.24  & 0.069 & 0.903 & $ 27.43 \pm 0.83 $  \\ \hline
${\rm (2) }$ & $0.2<P_E<1.0$ & 0.019 & 0.912
& 0.786 & 0.899 & 0.195 & 0.901 & $ 23.53 \pm 0.33 $  \\ \hline
${\rm (3) }$ & no cut        & 0.162 & --
& 0.582 & --   & 0.256 & --   & $ 27.05 \pm 0.27 $  \\ \hline
\multicolumn{9}{|c|}{Data at 189 GeV}\\ \hline
{\rm Sample} & b-tag prob. & $f\sp{(i)}_b$ & $C\sp{(i)}_b$ & $f\sp{(i)}_{uds}$
& $C\sp{(i)}_{uds}$ & $f\sp{(i)}_c$ & $C\sp{(i)}_c$ & $\langle n\rangle^{(i)}$ \\ \hline
${\rm (1) }$ & $P_E<0.00001$ & 0.899 & 0.912
& 0.016 & 1.15 & 0.085 & 0.919 & $ 27.75 \pm 0.48 $  \\ \hline
${\rm (2) }$ & $0.2<P_E<1.0$ & 0.016 & 0.896
& 0.789 & 0.893 & 0.195 & 0.913 & $ 23.93 \pm 0.24 $  \\ \hline
${\rm (3) }$ & no cut        & 0.161 & --
& 0.580 & --   & 0.259 & --   & $ 27.47 \pm 0.18 $  \\ \hline
\multicolumn{9}{|c|}{Data at 200 GeV}\\ \hline
{\rm Sample} & b-tag prob. & $f\sp{(i)}_b$ & $C\sp{(i)}_b$ & $f\sp{(i)}_{uds}$
& $C\sp{(i)}_{uds}$ & $f\sp{(i)}_c$ & $C\sp{(i)}_c$ & $\langle n\rangle^{(i)}$ \\ \hline
${\rm (1) }$ & $P_E<0.00001$ & 0.880 & 0.928
& 0.026 & 1.11 & 0.094 & 0.881 & $ 27.31 \pm 0.71 $  \\ \hline
${\rm (2) }$ & $0.2<P_E<1.0$ & 0.017 & 0.867
& 0.785 & 0.900 & 0.199 & 0.921 & $ 23.64 \pm 0.37 $  \\ \hline
${\rm (3) }$ & no cut        & 0.159 & --
& 0.579 & --   & 0.262 & --   & $ 27.52 \pm 0.29 $  \\ \hline
\end{tabular}
\end{center}
\caption{\label{puri}Mean 
multiplicities, $\langle n\rangle$,
in three event samples of different flavour content, $f_q$, and correction
factors $C_q$. The errors quoted on $\langle n\rangle$ are statistical only. The last
dataset contains only the data at 200~GeV from 1999.}
%selected by applying cuts on the b-tag probability, ${  P_E}$. }
\end{table}
                                                                         
In each of the three samples, the average multiplicity $\langle n\rangle$ is a linear
combination of the unknowns $\langle n\rangle_\bb$, $\langle n\rangle_{\ll}$
and $\langle n\rangle_\cc$.
One can thus formulate a set of three simultaneous equations to compute these
unknowns: 
\begin{eqnarray}
\langle n\rangle\sp{\rm (1)}
&=& f\sp{(1)}_b     C\sp{(1)}_b    \langle n\rangle_\bb
+ f\sp{(1)}_{uds} C\sp{(1)}_{uds} \langle n\rangle_{\ll}
+ f\sp{(1)}_c     C\sp{(1)}_c     \langle n\rangle_\cc\ \, ,\\ 
\langle n\rangle\sp{\rm (2)}
&=& f\sp{(2)}_b     C\sp{(2)}_b    \langle n\rangle_{\bb}
+ f\sp{(2)}_{uds} C\sp{(2)}_{uds} \langle n\rangle_{\ll}
+ f\sp{(2)}_c     C\sp{(2)}_c     \langle n\rangle_\cc\ \, ,\\
\langle n\rangle\sp{\rm (3)}
&=& f\sp{(3)}_b                    \langle n\rangle_\bb
+ f\sp{(3)}_{uds}                 \langle n\rangle_{\ll}
+ f\sp{(3)}_c                     \langle n\rangle_\cc\ .
\end{eqnarray}
Solving the above equations gave the following mean
charge multiplicities at 183 GeV:
\begin{eqnarray*}
\ & \langle n\rangle_\bb\rm{(183\,GeV)}    & = 29.79 \pm 1.11\ , \\
\ & \langle n\rangle_\cc\rm{(183\,GeV)}    & = 29.41 \pm 4.05\ , \\
\ & \langle n\rangle_{\lbl}\rm{(183\,GeV)} & = 25.25 \pm 1.35\ , \\
\ & \delta_{bl}\rm{(183\,GeV)} & = ~4.55 \pm 1.31\, ,
\end{eqnarray*}
with correlation coefficient of $-0.45$ between  $\langle n\rangle_\bb$ and 
$\langle n\rangle_{\lbl}$,
and at 189 GeV: 
\begin{eqnarray*}
\ & \langle n\rangle_\bb  \rm{(189\,GeV)}  & = 30.53 \pm 0.70\ , \\
\ & \langle n\rangle_\cc\rm{(189\,GeV)}    & = 28.63 \pm 2.81\ , \\
\ & \langle n\rangle_{\lbl}\rm{(189\,GeV)} & = 26.10 \pm 0.97\ , \\
\ & \delta_{bl}\rm{(189\,GeV)} & = ~4.43 \pm 0.85\, ,
\end{eqnarray*}
with correlation  coefficient of $-0.52$ between  $\langle n\rangle_\bb$ and  
$\langle n\rangle_{\lbl}$.

From the 1999 data, the results obtained for each energy 
are tabulated in Table \ref{eachen}.
The values were scaled to 200 GeV using JETSET and then a weighted 
average was calculated using the inverse of the square of the 
statistical error as weight. 
One obtains
\begin{eqnarray*}
\ & \langle n\rangle_\bb  \rm{(200\,GeV)}  & = 29.38 \pm 0.65\ , \\
\ & \langle n\rangle_\cc  \rm{(200\,GeV)}  & = 29.89 \pm 2.92\ , \\
\ & \langle n\rangle_{\lbl}\rm{(200\,GeV)} & = 25.99 \pm 1.03\ , \\
\ & \delta_{bl}\rm{(200\,GeV)} & = ~3.39 \pm 0.89\, ,
\end{eqnarray*}
with average correlation  coefficient of $-0.52$ between  $\langle n\rangle_\bb$ and
$\langle n\rangle_{\lbl}$.
The difference between the average of the values 
rescaled to 200 GeV and the average of the values without the scaling was
added in quadrature to the final systematic error. This difference is
anyway small (0.16 units for  $\langle n\rangle_\bb$ and less than 0.01 units
for $\delta_{bl}$).

The relatively large uncertainty of the
measured mean multiplicities for charm
stems from the inability of the $P_E$ variable
to extract a $c$-enriched sample of events.

It should be noted that the transition between particle and detector level 
measurements in equations (1) and (2) is done via multiplicative factors $C$ 
applied to the mean value of the distributions. The validity of this procedure 
requires that the simulation used to compute the $C$ values reproduces
the real data well; the $chi^2$/DF at centre of mass energies of 183, 189
and 200 GeV are respectively 0.81, 1.17 and 0.67 for the sample (1) and
0.93, 1.44 and 1.36 for the sample (2).

\begin{figure}
\mbox{\epsfxsize16cm\epsffile{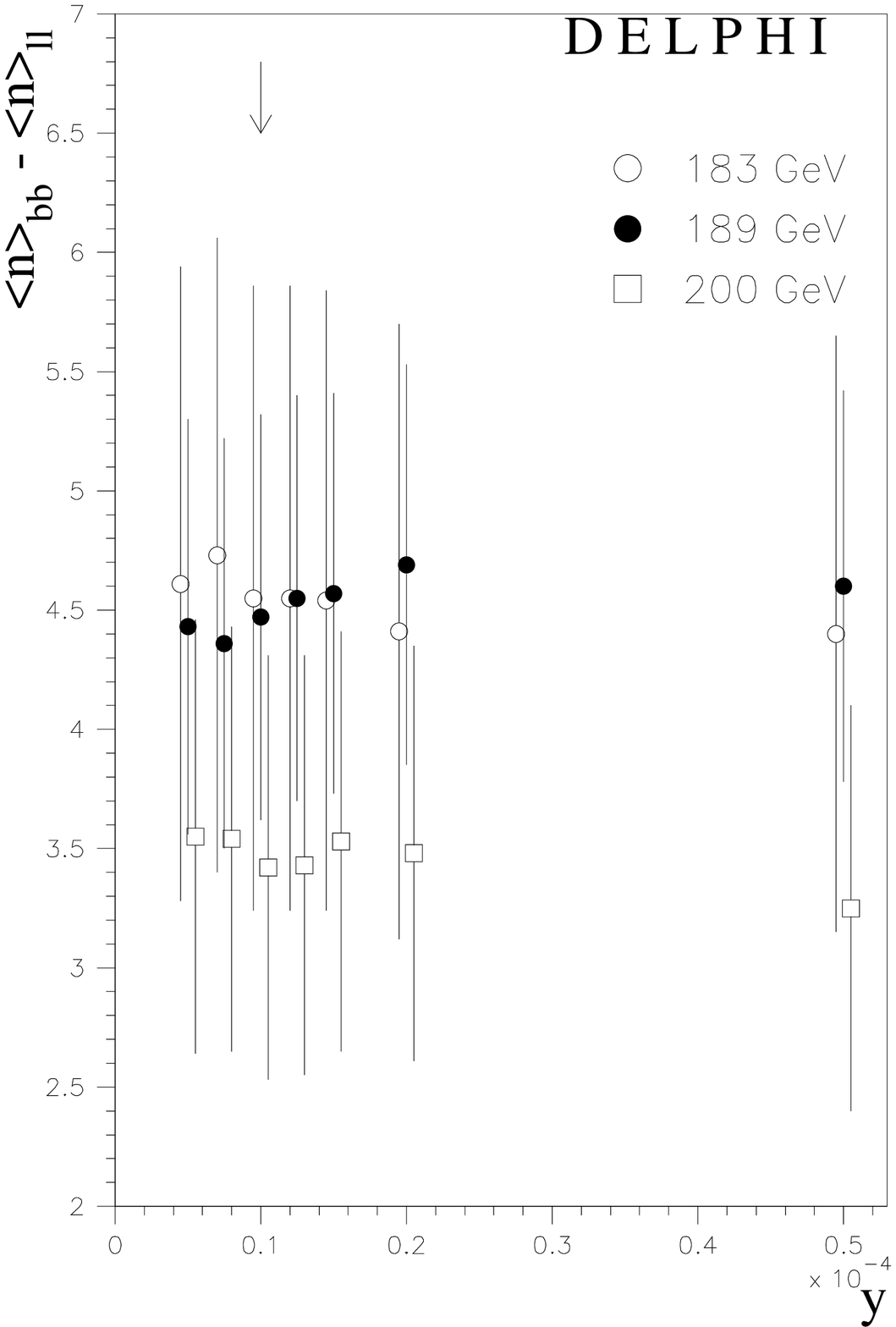}}
\caption[]{
Stability of $\delta_{bl}=\langle n\rangle_\bb-\langle n\rangle_{l\bar{l}}$ with respect to variations
of the 
cut on the  b-tagging variable, $y$. Notice that the errors 
in the plot are correlated (see text). The arrow indicates the
value used in the analyses.}\label{stab200}
\end{figure}

The analysis was repeated
with different cuts applied to the $b$-tag probability, $P_E$,
and the results for the $\delta_{bl}$ 
were found to be quite stable (see Figure~\ref{stab200}).
A systematic error was evaluated as
half of the difference between the greatest and the smallest 
multiplicity values obtained from
varying the cut on $P_E$ from $0.5\times10^{-5}$ to $1.5\times10^{-5}$.
\begin{table}[t]
\begin{center}
\begin{tabular}{|l|c|c|c|c|}\hline
$E_{cm}$ & $\langle n\rangle_\bb$  & $\langle n\rangle_\cc$  & $\langle n\rangle_\lbl$ & $\delta_{bl}$ \\ \hline
192 GeV & $27.57\pm 1.56$ & $30.63\pm7.70$ & $25.54\pm2.75$ & $2.03\pm2.36$ \\
196 GeV & $29.58\pm 0.97$ & $26.75\pm4.45$ & $27.12\pm1.58$ & $2.46\pm1.37$ \\
200 GeV & $29.55\pm 1.06$ & $32.42\pm4.43$ & $24.75\pm1.54$ & $4.79\pm1.34$ \\ \hline
\end{tabular}
\end{center}
\caption{\label{eachen}Multiplicities measured for each energy during 1999.}
\end{table}

The uncertainty due to the event selection in sample (2) was investigated by
repeating the analysis after variation of the narrow jet broadening cut,
from 0.05 to 0.08. 
Half of the differences between the greatest and the smallest multiplicities
were added in quadrature to the systematic error previously calculated.
The propagated systematic error in the total multiplicity 
in equation (3) from \cite{noi}
was also added in quadrature to the systematic error.
%Finally, 
Uncertainties arising from the modelling of short-lived particles
in the simulation were considered. 
The main physics sources of these uncertainties
come from the assumed lifetime of B-hadrons
($\tau_B = 1.564 \pm 0.014$ ps) \cite{pdg},
and the D$\sp{+},$ D$\sp{0}$ lifetimes and production rates
\cite{pdg}. 
Also a variation in the modelling of the $b$ fragmentation
was investigated, by allowing the average fractional energy of a $B$ hadron
to vary by 1.5\%.
The same relative uncertainty was assumed as in \cite{delphi}. 

The effect of a variation of 1\% 
in the fraction $R_b$
of $\bb$ events and of 3\% 
in the fraction $R_c$ of $\cc$ events was found to be negligible.
Since the multiplicity difference, $\delta_{bl}$, was found to be
independent of energy within errors, the effect of the
modelling of the initial state radiation is also expected to be negligible.

The contributions to the systematic error are summarized in Table \ref{sys}.
\begin{table}[htbp]
\begin{center}
\begin{tabular}{|l|c|c|c|c|c|c|}\hline
 & \multicolumn{2}{|c|}{183 GeV} & \multicolumn{2}{|c|}{189 GeV}  & \multicolumn{2}{|c|}{200 GeV} \\
Source & $\langle n\rangle_{\bb}$ & $\delta_{bl}$ & $\langle n\rangle_{\bb}$ &
$\delta_{bl}$  & $\langle n\rangle_{\bb}$ & $\delta_{bl}$ \\ \hline
$b$-tag probability cut   & 0.14 & 0.10 & 0.16 & 0.11  & 0.25 & 0.17\\
Narrow jet broadening cut & 0.06 & 0.32 & 0.05 & 0.18  & 0.15 & 0.63\\
Modelling in the simulation & 0.10 & 0.33 & 0.10 & 0.32  & 0.09 & 0.23\\
$E_{cm}$ rescaling & -- & -- & -- & --  & 0.16 & 0.00 \\
Systematic error on $\langle n\rangle^{(3)}$ & 0.21 & 0.56 & 0.27 & 0.47 & 0.36 & 0.74 \\ \hline
TOTAL                     & 0.28 & 0.73 & 0.34 & 0.61 & 0.50 & 1.01 \\ \hline
\end{tabular}
\end{center}
\caption{\label{sys}Contributions to the systematic errors on $\langle n\rangle_{\bb}$ 
and $\delta_{bl}$.}
\end{table}

The final mean values of
the event multiplicity in b events are
$\langle ~n~\rangle_\bb\rm{(183\,GeV)}$      = $29.79 \pm 1.11(stat) \pm 0.28(syst)$, 
$\langle ~n~\rangle_\bb\rm{(189\,GeV)}$      = $30.53 \pm 0.70(stat) \pm 0.34(syst)$, and
$\langle ~n~\rangle_\bb\rm{(200\,GeV)}$      = $29.38 \pm 0.65(stat) \pm 0.50(syst)$.
The multiplicity 
difference between $\bb$ and light quark-antiquark events
measured at the different energies is:
\begin{eqnarray}\label{e:mulc}
\ & \delta_{bl}\rm{(183\,GeV)} & = ~4.55 \pm 1.31(stat) \pm 0.73(syst)\ , \\
\ & \delta_{bl}\rm{(189\,GeV)} & = ~4.43 \pm 0.85(stat) \pm 0.61(syst)\ , \\
\ & \delta_{bl}\rm{(200\,GeV)} & = ~3.39 \pm 0.89(stat) \pm 1.01(syst)\ .
\end{eqnarray}
These values include the products of K$^0_S$ and $\Lambda$ decays.
The uncertainties on the modelling of the detector largely cancel out
in the difference.

Our results on $\delta_{bl}$ are plotted in Figure \ref{nice} and compared with
previous results in the literature.  

\section{Comparison with Models and QCD Predictions}

{\bf{Flavour-Independent Fragmentation ---}} 
In a model in which the hadronization is independent of the 
mass of the quarks,
one can assume that the non-leading multiplicity
in an event,
i.e., the light quark multiplicity which accompanies 
the decay products of the primary hadrons,
is governed by the effective energy available
to the fragmentation system following the
production of the primary hadrons \cite{kisselev}.  
One can thus write:
\begin{eqnarray}
\delta_{bl}(E_{cm}) & = & 2\langle n^{(decay)}_B\rangle  + 
 \int_0^1  dx_B f_{E_{cm}}(x_B) \int_0^1 dx_{\bar{B}} f_{E_{cm}}(x_{\bar{B}}) 
\, \,
 n_{l\bar{l}}\left( \left(1-\frac{x_B+x_{\bar{B}}}{2} \right) 
 E_{cm}\right)\nonumber \\
 & - & n_{l\bar{l}}(E_{cm}) \label{strafi}\, ,
%\nonumber \\
% & \simeq & 2<n^{(decay)}_B>  + 
% n_{l\bar{l}}\left( \left(1-<x_B>\right) 
% E_{cm}\right)\nonumber \\
%  -  n_{l\bar{l}}(E_{cm})
\end{eqnarray}
where 
$\langle n^{(decay)}_B\rangle$ is the average number of charged 
particles coming from the
decay of a $B$ hadron,
$x_B$ ($x_{\bar{B}}$) is the fraction of the beam energy taken by
the $B$ ($\bar{B}$) hadron, and $f_{E_{cm}}(x_B)$ is the $b$ fragmentation
function.

We assumed $2\langle n^{(decay)}_B\rangle = 11.0 \pm 0.2$ \cite{schumm}, 
consistent with the average $\langle n^{(decay)}_B\rangle = 5.7 \pm 0.3$
measured at LEP \cite{vietri}.
For $f_{E_{cm}}(x_B)$, we assumed a Peterson function 
%of average $0.70 \pm 0.02$
with hardness parameter $\epsilon_p=0.0047^{+0.0010}_{-0.0008}$
\cite{pdg}, evolving with energy as in~\cite{lund} to take into account the
effects of scaling violations.
The value of $n_{\ll}(E)$ was computed from the fit
to a perturbative 
QCD formula \cite{webber} including the resummation of leading
(LLA) and next-to-leading (NLLA) corrections,
which reproduces well the 
measured charged multiplicities \cite{noi}, with appropriate 
corrections to remove the effect of heavy quarks \cite{dea} and leading 
particles.

\begin{figure}
\mbox{\epsfxsize16cm\epsffile{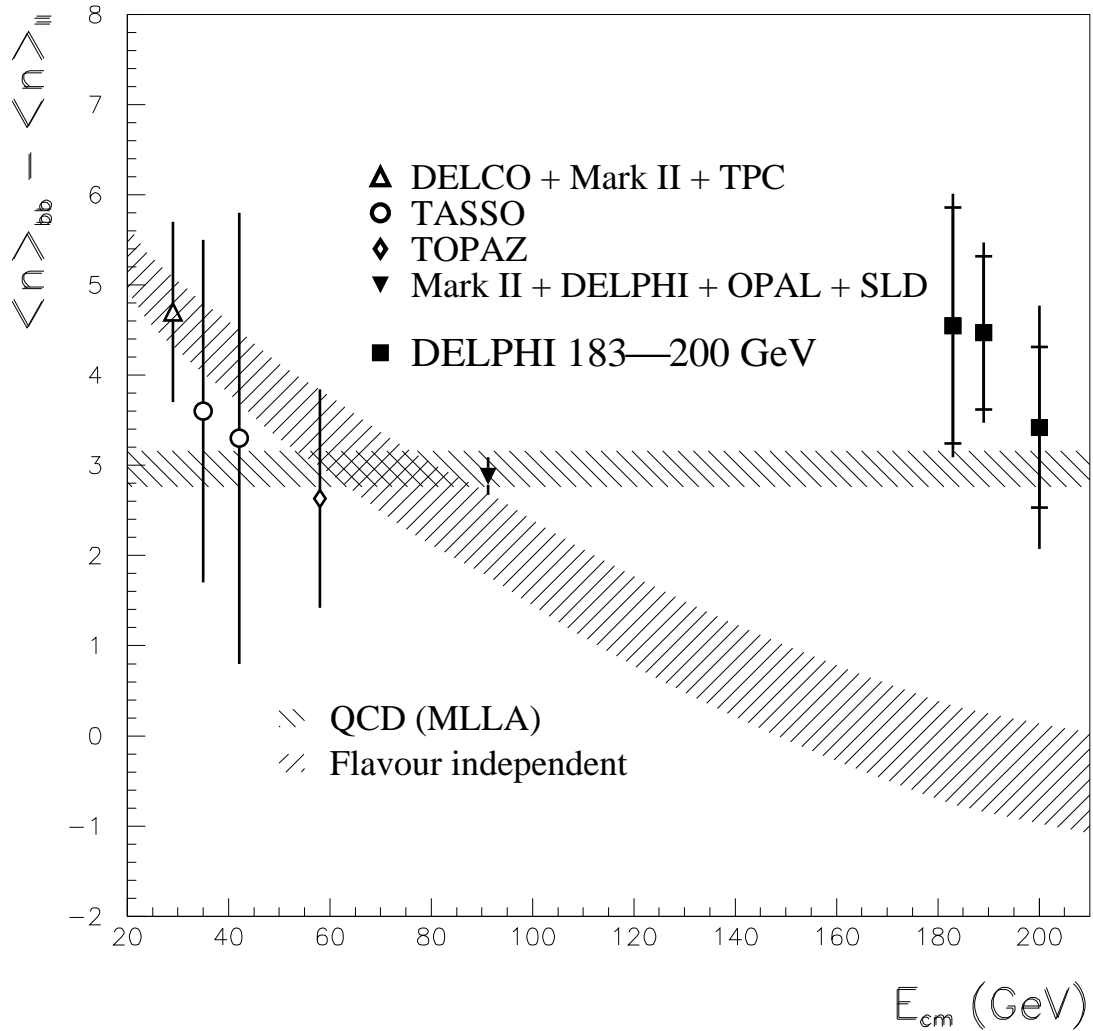}}
\caption[]{
The present measurement of $\delta_{bl}$ 
compared to previous measurements as a function of the centre-of-mass energy,
to the QCD prediction (taken as the average of the values
up to the Z included, see the text), and to the expectation from 
flavour-independent 
fragmentation. The inner error bars represent the statistical error;
the full bars show the sum in quadrature of the statistical and 
systematic errors.}\label{nice}
\end{figure}

The prediction of the model in which the hadronization
is independent of the quark mass is plotted in Figure \ref{nice}.
The reason for the drop with collision energy is
that the heavy quark system carries away a large fraction of
the available energy, approximately (i.e., neglecting scaling violations)
linear with $\sqrt{s}$, 
while the multiplicity growth with $\sqrt{s}$ is less than linear.
There are several variations of this
model in the literature, 
leading to slightly different predictions
(see \cite{vietri} and references therein).
The result from substituting in Eq.~(\ref{strafi})
$n_{l\bar{l}}\left( \left(1-\frac{x_B+x_{\bar{B}}}{2} \right) E_{cm}\right)$
with
$n_{l\bar{l}}\left( E_{cm}\sqrt{(1-x_B)(1-x_{\bar{B}})} \right)$ as 
in~\cite{opal}, or approximating the Peterson fragmentation function with a 
Dirac delta function at $\langle x_B\rangle$, are within the errors. Also by using for
$n_{l\bar{l}}$ the expression in~\cite{opal} one stays within the band
in Figure~\ref{nice}. The prediction as plotted in  Figure~\ref{nice}
agrees with the one calculated in \cite{khoze}.

{\bf{QCD Calculation ---}} 
The large mass of the $b$ quark,
in comparison to the scale of the strong interaction,
$\Lambda \simeq 0.2$ GeV, results in a natural 
cut off for the emission of gluon bremsstrahlung.
Furthermore, where the c.m. energy 
greatly exceeds the scale of the $b$ quark mass, 
the inclusive spectrum of heavy quark production is
expected to be well described by perturbative QCD
in the Modified Leading Logarithmic Approximation (MLLA, \cite{book}).

The value of $\delta_{bl}$
has been calculated in perturbative QCD\cite{schumm,petrov}:
\begin{equation}\label{qcd}
\delta_{bl} =  
2\langle n^{(decay)}_B\rangle
- \langle n_{l\bar{l}}\rangle(\sqrt{s}=e^{1/2}m_b) +  
O(\alpha_s(m_b))\langle n_{l\bar{l}}\rangle(\sqrt{s}=m_b) \, .
\end{equation}
The reason for the appearance of the $e^{1/2}$ factor in the above 
expression is discussed in detail in \cite{petrov}.
The calculation of the actual value of
$\delta_{bl}$ in \cite{schumm}
on the basis of the first two terms in (\ref{qcd}) gives
a value of $5.5 \pm 0.8$. 
A different calculation of $\delta_{bl}$
gives 3.68 \cite{petrov}. 
These two calculations assume $m_b=5$ GeV/$c^2$ and
$m_b=4.8$ GeV/$c^2$ respectively, and different parametrizations
for the function $\langle n_{l\bar{l}}\rangle(\sqrt{s})$. The dependence of
the perturbative part in Eq. (\ref{qcd}) on $m_b$ is such that
moving the $m_b$ value from 5 GeV/$c^2$ to 4 GeV/$c^2$
induces a change of +0.6 units of multiplicity.

The difference of the results in \cite{schumm} and in \cite{petrov}
demonstrates the importance of the contribution
proportional to $\alpha_s(m_b)$.
A less restrictive condition is the calculation of upper limits:
an upper limit $\delta_{bl} < 4.1$ is given in \cite{petrov}, 
based on the maximization of the nonperturbative term; 
$\delta_{bl} < 4$ is obtained from phenomenological arguments
in Ref. \cite{deus}.

Although the presence of the last term in the equation limits the accuracy in
the calculation of $\delta_{bl}$, 
QCD tells that $\delta_{bl}$ is fairly 
independent of $E_{cm}$.
In this article  the
average of the experimental values of $\delta_{bl}$ up to $m_Z$
included, $\langle \delta_{bl}\rangle = 2.96 \pm 0.20$ (dominated by the LEP 1 data), 
is taken as
the high energy prediction from QCD. The accuracy of the measurement at the Z
is thus used to constrain the theoretical prediction.

Our  measurement of $\delta_{bl}$, as seen in Figure \ref{nice}, 
is consistent with the
prediction of energy independence based on perturbative QCD, and
more than three standard deviations larger than
predicted by the naive model presented in the beginning of this
section.

\section{Conclusions}

The difference $\delta_{bl}$ between the average charged particle multiplicity 
$\langle n\rangle_\bb$ in $e^+e^- \rightarrow b\bar{b}$ events
and the multiplicity in generic light quark $l = u,d,s$ events
has been measured at centre-of-mass energies of 183, 189 and 200 GeV:
\begin{eqnarray*}
\delta_{bl}\rm{(183\,GeV)} &=& 4.55 \pm 1.31 (stat) \pm 0.73 (syst)\\
\delta_{bl}\rm{(189\,GeV)} &=& 4.43 \pm 0.85 (stat) \pm 0.61 (syst)\\
\delta_{bl}\rm{(200\,GeV)} &=& 3.39 \pm 0.89(stat) \pm 1.01(syst)\, .
\end{eqnarray*}
This difference is in agreement with QCD predictions, while it is 
inconsistent with calculations assuming that 
the multiplicity accompanying the decay of a heavy quark is independent of the 
mass of the quark itself. 

\subsection*{Acknowledgements}
We are grateful to Jorge Dias
de Deus, Vladimir Petrov, Alexander Kisselev, 
Valery Khoze and Torbj\"orn Sj\"ostrand for
useful discussions.\\ 
%         Modified on 04-06-1999 by dimartino
%-------------------------------------------------------------------
%\subsection*{Acknowledgements}
%\vskip 3 mm
 We are greatly indebted to our technical 
collaborators, to the members of the CERN-SL Division for the excellent 
performance of the LEP collider, and to the funding agencies for their
support in building and operating the DELPHI detector.\\
We acknowledge in particular the support of \\
Austrian Federal Ministry of Science and Traffics, GZ 616.364/2-III/2a/98, \\
FNRS--FWO, Belgium,  \\
FINEP, CNPq, CAPES, FUJB and FAPERJ, Brazil, \\
Czech Ministry of Industry and Trade, GA CR 202/96/0450 and GA AVCR A1010521,\\
Danish Natural Research Council, \\
Commission of the European Communities (DG XII), \\
Direction des Sciences de la Mati$\grave{\mbox{\rm e}}$re, CEA, France, \\
Bundesministerium f$\ddot{\mbox{\rm u}}$r Bildung, Wissenschaft, Forschung 
und Technologie, Germany,\\
General Secretariat for Research and Technology, Greece, \\
National Science Foundation (NWO) and Foundation for Research on Matter (FOM),
The Netherlands, \\
Norwegian Research Council,  \\
State Committee for Scientific Research, Poland, 2P03B06015, 2P03B1116 and
SPUB/P03/178/98, \\
JNICT--Junta Nacional de Investiga\c{c}\~{a}o Cient\'{\i}fica 
e Tecnol$\acute{\mbox{\rm o}}$gica, Portugal, \\
Vedecka grantova agentura MS SR, Slovakia, Nr. 95/5195/134, \\
Ministry of Science and Technology of the Republic of Slovenia, \\
CICYT, Spain, AEN96--1661 and AEN96-1681,  \\
The Swedish Natural Science Research Council,      \\
Particle Physics and Astronomy Research Council, UK, \\
Department of Energy, USA, DE--FG02--94ER40817. \\
%=========================================================================%

\newpage

\newpage

\end{document}